\documentclass[a4paper,11pt]{article}
\usepackage{pos}
\usepackage{braket}
\usepackage{amsmath}
\usepackage{amssymb}
\usepackage{multicol}
\usepackage{slashed}
\usepackage{mathtools}
\usepackage{float}
\usepackage{dsfont}
\newcommand{\Tr}{{\rm Tr}}

\usepackage[utf8]{inputenc}\DeclareUnicodeCharacter{2212}{-}
\usepackage[normalem]{ulem}

\title{The glueball spectrum with $N_f=4$ light fermions}

\newcommand{\be}{\begin{equation}}
\newcommand{\ee}{\end{equation}}
\newcommand{\bea}{\begin{eqnarray}}
\newcommand{\eea}{\end{eqnarray}}

\author*[a,b]{Andreas Athenodorou}
\author[a]{Jacob Finkenrath}
\author[c]{Adam Lantos}
\author[d,e]{Michael~Teper}

\affiliation[a]{Computation-based Science and Technology Research Center, The Cyprus Institute, Cyprus}
\affiliation[b]{Dipartimento di Fisica, Universit\'a di Pisa and INFN, Sezione di Pisa, Largo Pontecorvo 3, 56127 Pisa, Italy}
\affiliation[c]{Cyprus University of Technology, Archiepiskopou Kyprianou 30, Limassol 3036}
\affiliation[d]{Rudolf Peierls Centre for Theoretical Physics, University of Oxford, Parks Road, Oxford OX1 3PU, UK}
\affiliation[e]{All Souls College, University of Oxford, High Street, Oxford OX1 4AL, UK}

\emailAdd{a.athenodorou@cyi.ac.cy}
\emailAdd{j.finkenrath@cyi.ac.cy}
\emailAdd{mike.teper@physics.ox.ac.uk}

\abstract{We investigate the glueball spectrum for $N_f=4$ fermions corresponding to low pion masses of $m_\pi \sim 250$MeV. We do so by making use of configurations produced with maximally twisted fermions within the framework of the Extended Twisted Mass Collaboration (ETMC). We extract states that belong to irreducible representations of the octahedral group of rotations $R$ in combination with the quantum numbers of charge conjugation $C$ and parity $P$, i.e. $R^{PC}$. We implement the Generalized Eigenvalue Problem (GEVP) using a basis consisting only of gluonic operators. The purpose of this work is to investigate the effect of light dynamical quarks on the glueball spectrum and how this compares to the statistically more accurate spectrum of the pure gauge theory. We employed large ensembles of the order of ${\sim {~\cal O}}(10 {\rm K})$ configurations for each of three different lattice spacings. Our results demonstrate that in the scalar channel $A_1^{++}$ we obtain an additional, lightest state due to the inclusion of light dynamical quarks while the next two states are consistent with the lightest two states in the pure gauge theory. By contrast the mass of the lightest tensor glueball $J^{PC}=2^{++}$ appears to be insensitive to the inclusion of sea quarks, as is the mass of the lightest pseudoscalar. In addition we perform an investigation of the low lying spectrum of the representation $A_1^{++}$ for $N_f=2+1+1$ twisted mass quarks with low masses and demonstrate that the extra lowest state depends strongly on the pion mass. This suggests that the ground state of the scalar glueball has a large quark content, possibly representing the decay of a glueball to two pions.}

\FullConference{%
  The 39th International Symposium on Lattice Field Theory (Lattice2022),\\
  8-13 August, 2022 \\
  Bonn, Germany 
}

\begin{document}
\maketitle

\section{Introduction}
\label{sec:introduction}
The extraction of the spectrum of glueballs in full QCD at physical quark masses is still an open question which requires a coordinated effort from the Lattice community. Given that a number of experiments such as PANDA~\cite{Parganlija:2013xsa} and BESIII~\cite{Asner:2008nq} are currently looking for such states,  the extraction of the associated spectrum and the understanding of how light dynamical quarks affect the spectrum of glueballs is a timely question. Recent reviews on searches for glueballs can be found in the Lattice 2022 plenary presentation by Davide Vadacchino~\cite{vadacchino_davide_2022_7338133} as well as in Ref.~\cite{Klempt:2022ipu}. Glueball states have large fluctuations and possibly a broad resonance width and so the precise determination of their spectrum requires high statistics of the order of ${{\cal O}} (10-100{\rm K})$ configurations. Hadron physics performed using current Lattice methodology and machinery, on the other hand, makes use of sets of typically a few hundred configurations. Nevertheless, for a number of measurements on the lattice such as the extraction of renormalisation factors, large sets of configurations are produced and can be, thus, used for exploratory investigations of the glueball spectrum. This work results from the availability of accessing such large sets of configurations.  

In this work we investigate the effect of light quarks on the spectrum of glueballs. The spectrum of QCD with light quarks is of interest for both phenomenlogical and theoretical reasons. On the phenomenological side it would provide important information regarding the mixing of ${q {\bar q}}$ states with glueball states which is expected to be modest by the OZI rule~\cite{OKUBO1963165,10.1143/PTPS.37.21}; this could mean that in practice states with quark content could actually be undetectable in glueball spectroscopy or they appear with extremely small overlaps. For this purpose we use configurations produced with $N_f = 4$ light twisted mass quarks with masses corresponding to $m_{\pi} \sim 250$~MeV, i.e. not very far from the physical quark masses. We extract the glueball spectrum and then compare it with the one extracted using pure gauge $SU(3)$ configurations~\cite{Athenodorou:2020ani, Athenodorou:2021qvs}. We expect that such a set up will enhance the contribution of light dynamical quarks to the spectrum of glueballs. For massive dynamical quarks we expect, from decoupling arguments, that the glueball spectrum becomes similar to the spectrum of the pure gauge theory. The important question which arises here is what happens if one includes dynamical fermions with low masses close to physical values.

Our investigation for $N_f=4$ QCD with light fermions reveals, at least in the scalar channel ($R^{PC} = A_1^{++}$), an additional state which is also the lightest state. Possibly this state involves some mixing of $q {\bar q}$ states. To understand what occurs in the low-lying spectrum of the scalar channel and whether this additional state includes $q {\bar q}$ states we turn to $N_f=2+1+1$ configurations. Within this setup we extract the ground state energy for the scalar channel which appears to exhibit a strong dependence on the pion mass. Overall, our main findings can be summarized in the following points. (1) In the scalar channel we obtain an additional state when we introduce light dynamical quarks, and this is the lightest state with the next two states having masses consistent with the lightest two glueballs in the pure gauge theory. (2) The tensor glueball mass appears to be insensitive to the presence of light dynamical quarks. (3) The pseudoscalar glueball mass is affected only slightly by the appearance of light dynamical quarks and is close to the mass of the tensor glueball. \newpage

Following the structure of the actual presentation~\cite{andreas_athenodorou_2022_6982917} this paper is organised as follows. In Section~\ref{sec:simulation_details} we present the lattice set up used to produce the configurations with $N_f=4$ and $N_f=2+1+1$ twisted mass fermions as well as those with the pure gauge action. Then, in Section~\ref{sec:glueball_masses} we explain briefly how one can extract the spectrum of glueballs in Lattice QCD by making use of the Generalized Eigenvalue Problem (GEVP) method. Following that, in Section~\ref{sec:topological_charge_and_scale_setting}, we describe the calculation of the topological charge which is used as a measure of the ergodicity of the system. Furthermore we explain how we evaluate the energy scale $t_0$ using the smoothing scheme of the gradient flow. Subsequently, we move to the presentation of the results by focussing on the scalar channel $R^{PC}=A_1^{++}$, the tensor glueball whose components are split between the $R^{PC}=E^{++}$ and the $T_2^{++}$ channels, as well as the pseudoscalar glueball obtained in the $R^{PC}=A_1^{-+}$ channel. In Section~\ref{sec:Nf211} we focus our discussion on the ground state obtained in the scalar channel for $N_f=2+1+1$ which appears to include $q {\bar q}$ states. Finally, in Section~\ref{sec:conclusions} we present our conclusions.  

\section{Simulation Details}
\label{sec:simulation_details}
\vspace{-0.25cm}
We use gauge ensembles of clover improved twisted mass fermions produced  with \(4\) degenerate light flavours ($N_f=4$) at two different lattice spacings as well as two ensembles with \(2\) degenerate light flavours and a strange and a charm quark ($N_f=2+1+1$). All the configurations with dynamical quarks have been generated within the context of the Extended Twisted Mass collaboration.


We use the Iwasaki improved gauge action~\cite{Iwasaki:1984cj,Iwasaki:1983gx,Weisz:1982zw}, which is given by the expression
\be
S_G = \frac{\beta}{3} \sum_x \left(c_0\sum_{ \substack{\mu,\nu=1 \\ \mu<\nu} }^4\left[ 1- \Re\Tr\left( U_{x,\mu\nu}^{1\times1}\right)\right] + c_1 \sum_{ \substack{\mu,\nu=1 \\ \mu\neq\nu} }^4 \left[ 1- \Re\Tr\left(U_{x,\mu\nu}^{1\times2}\right)\right]\right)\,,
\label{eq:g_action}
\ee
where \(\beta=6/g^2\), \(U^{1\times1}\) is a plaquette and \(U^{1\times2}\) is a rectangular Wilson loop. The Symanzik coefficients are \(c_0=3.648\) and \(c_1= (1-c_0)/8\). The fermionic action is given
 \cite{Frezzotti:2000nk,Frezzotti:2003ni} by:
\be
S_F^{l} = a^4\sum_x \bar{\chi}^{(l)}(x)\left( D_W[U] + \frac{i}{4} c_{SW}\sigma^{\mu\nu}\mathcal{F}^{\mu\nu}[U] + m_{0,l} + i \mu_l\gamma_5\tau^3 \right)\chi^{(l)}(x)\,.
\label{eq:fl_action}
\ee
In the equation above, \(\chi^{(l)}\) is the field representing the  quark doublets, expressed in the twisted basis, \(m_{0,l}\) and \(\mu_l\) are respectively the untwisted and twisted  mass parameters, \(\tau^3\) is the third Pauli matrix acting in flavor space and \(D_W\) is the massless Wilson-Dirac operator. The clover term \(\propto \sigma^{\mu\nu}\mathcal{F}^{\mu\nu}\) is included in the action to suppress cut-off effects reducing the difference between the mass of the charged and neutral pions~\cite{Alexandrou:2018egz}.

Turning now to the case of $N_f=2+1+1$, the additional strange and charm quarks are included as a non-degenerate twisted doublet \(\chi^{(h)}=(s,c)^t\), with the  action~\cite{Frezzotti:2003xj}
\be
S_F^{h} = a^4\sum_x \bar{\chi}^{(h)}(x)\left( D_W[U] + \frac{i}{4} c_{SW}\sigma^{\mu\nu}\mathcal{F}^{\mu\nu}[U] + m_{0,h} - \mu_{\delta}\tau^1 + i \mu_{\sigma}\gamma_5\tau^3 \right)\chi^{(h)}(x)\,,
\label{eq:fh_action}
\ee
 where \(m_{0,h}\) is the bare untwisted quark mass for the heavy doublet, \(\mu_{\delta}\) the bare twisted mass along the \(\tau^1\) direction and \(\mu_{\sigma}\) the mass splitting in the \(\tau^3\) direction.

The partial conserved axial current (PCAC) mass is tuned to zero in order to achieve maximal twist. This ensures automatic \(\mathcal{O}(a)\) improvement for the expectation values of the observables of interest~\cite{Frezzotti:2005gi}. The simulation parameters of the gauge ensembles with $N_f=4$ as well as $N_f=2+1+1$ quarks are given in Tables~\ref{tab:params_sim_Nf4} and \ref{tab:params_sim_Nf2+1+1} respectively.



\begin{table}[H]
    \centering
    \begin{tabular}{l|c|l|c|c|c|c}
        \hline
  & $\beta$ & $\quad c_{SW}$ & $\mu_l$ & $L$ & $a m_{PS}$ & $t_0/a^2$ \\
        \hline
\texttt{cB4.06.16} & $1.778$ & $1.69$ & $0.006$ & 16 & $0.2652(53)$ & $4.947(62)$  \\
\texttt{cB4.06.24} & $1.778$ & $1.69$ & $0.006$ & 24 & $0.1580(\phantom{0}8)$ & $4.667(17)$  \\
\texttt{cC4.05.24} & $1.836$ & $1.6452$ & $0.005$ & 24 & $0.1546(20)$ & $6.422(48)$  \\
\hline
\end{tabular}
\captionof{table}{ Simulation parameters of the $N_f=4$ gauge ensembles~\cite{Alexandrou:2018egz,Alexandrou:2018sjm} used in this work.}
    \label{tab:params_sim_Nf4}
\end{table}
\begin{table}[H]
    \vspace{-0.5cm}
    \centering
    \begin{tabular}{l|c|l|c|c|c|c}
        \hline
  & $\beta$ & $\quad c_{SW}$ & $\mu_l$ & $L$ & $a m_{PS}$ & $t_0/a^2$ \\
        \hline
\texttt{cA211.53.24} & $1.726$ & $1.74$ & $0.0053$ & 24 & $0.1661(\phantom{0}4)$ & $2.342(\phantom{0}6)$  \\
\texttt{cA211.25.32} & $1.726$ & $1.74$ & $0.0025$ & 32 & $0.1253(\phantom{0}1)$ & $2.392(\phantom{0}4)$  \\
\hline
\end{tabular}
\captionof{table}{ Simulation parameters of the $N_f=2+1+1$ gauge ensembles ~\cite{Alexandrou:2018egz,Alexandrou:2018sjm} used in this work.}
    \label{tab:params_sim_Nf2+1+1}
    \vspace{-0.25cm}
\end{table}
For purposes of comparison with the $SU(3)$ pure gauge theory, we also simulate that theory using the standard Wilson action which is given by Eq.~\ref{eq:g_action} with $c_0=1$ and $c_1=0$. The simulation algorithm combines standard heat-bath and over-relaxation steps in the ratio 1:4; these are implemented by updating $SU(2)$ subgroups using the Cabibbo-Marinari algorithm~\cite{Cabibbo:1982zn}. The parameters of the pure gauge runs are provided in Table~\ref{tab:params_sim_pure_gauge}.
\begin{table}[H]
    \centering
    \begin{tabular}{c|c|c|c|c}
        \hline
 $\beta$ & $L$ & $\sqrt{\sigma}$ & $a m_{G}$ & $t_0/a^2$ \\
        \hline
$6.222$ & $30$ & $0.1533(6)$ & $0.499(6)$ & $6.422(48)$ \\    
$6.135$ & $26$ & $0.1750(9)$ & $0.578(4)$  & $4.947(62)$  \\
$6.117$ & $26$ & $0.1781(12)$ & $0.585(8)$ & $4.667(17)$  \\
\hline
\end{tabular}
\captionof{table}{Simulation parameters for the $SU(3)$ pure gauge ensembles used in this work. The last column represents the values of $t_0/a^2$ for $N_f=4$ simulations corresponding to the given values of $\beta$ via a cubic spline interpolation of data from Refs.~\cite{Luscher:2010iy,Francis:2015lha,Luscher:2011kk,Ce:2015qha}}
    \label{tab:params_sim_pure_gauge}
\end{table}
The idea behind choosing the particular parameters is based upon matching the gradient flow times $t_0/a^2$ between the $N_f=4$ QCD runs and the pure gauge runs. Since $N_f=4$ QCD configurations had already been produced, we calculate the corresponding values for $t_0/a^2$ and we subsequently match them to the right value of $\beta$ for the pure gauge theory using interpolations based on data from Refs~\cite{Luscher:2010iy,Francis:2015lha,Luscher:2011kk,Ce:2015qha}. The motivation for this procedure is discussed in Section~\ref{sec:results}.

\section{Calculation of glueball masses}
\label{sec:glueball_masses}

\vspace{-0.25cm}
Glueballs are colour singlet states. Masses of colour singlet states can be calculated using the standard decomposition of a Euclidean correlator of an operator $\phi(t)$ onto physical states in terms of the energy eigenstates of the Hamiltonian of the system $H$:
\begin{eqnarray}
\langle \phi^\dagger(t=an_t)\phi(0) \rangle
 = 
\langle \phi^\dagger e^{-Han_t} \phi \rangle
= \sum_i |c_i|^2 e^{-aE_in_t} 
\stackrel{t\to \infty}{=} 
|c_0|^2 e^{-aE_0n_t}\,.
\label{extract_mass}
\end{eqnarray}
Here the energy levels are ordered, $E_{i+1}\geq E_i$, with  $E_0$ that of the ground state. The only states that contribute in the above summation are those that have non zero overlaps i.e. $c_i = \langle {\rm vac} | \phi^\dagger | i \rangle \neq 0$. We therefore need to match the quantum numbers of the operator $\phi$ to those of the state we are interested in. 

The extraction of the ground state relies on how good the overlap is onto this state and how fast in $t$ we obtain the exponential decay according to Eq.~(\ref{extract_mass}). The overlap can be maximized by building operator(s) which "capture" the right properties of the state, in other words by projecting onto the physical length scales of the relevant state as well as onto the right quantum numbers. To this end we employ the GEVP~\cite{Luscher:1984is,Luscher:1990ck} applied to a basis of operators built by several lattice loops in different blocking levels \cite{Lucini:2004my,Teper:1987wt}. This reduces the contamination of excited states onto the ground state and maximizes the overlap of the operators onto the physical length scales.

\begin{figure}[h]
    \centering
    \includegraphics[height=3.75cm]{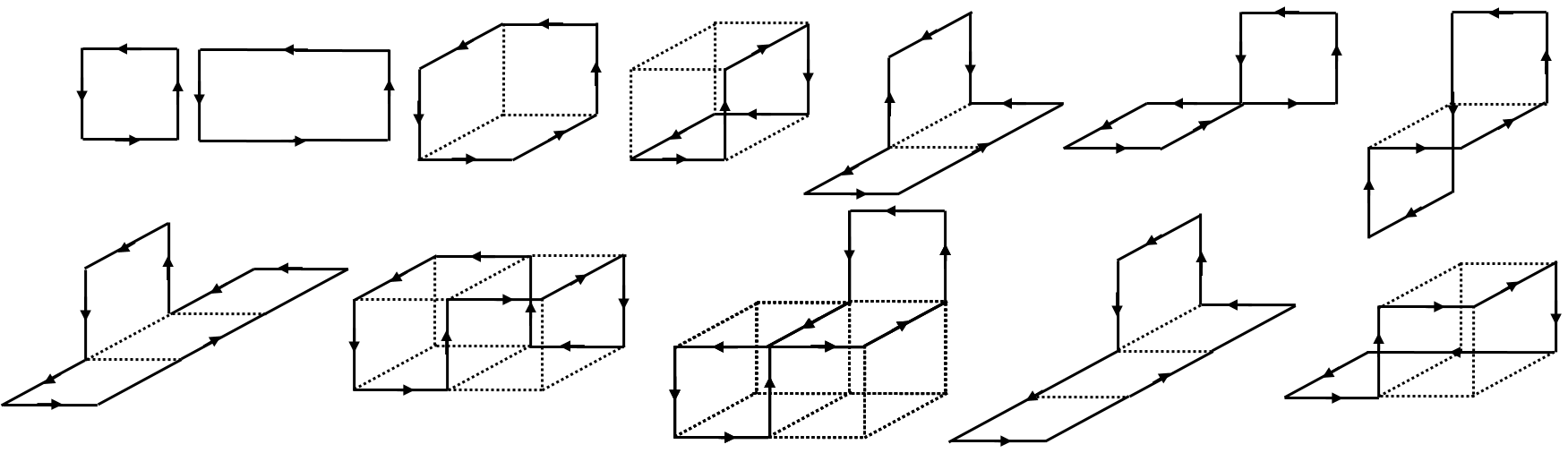}
    \caption{All the different closed loops used for the construction of the glueball operators.}
    \label{fig:glueball_operators}
\end{figure}

The glueballs are color singlets and, thus, an operator projecting onto a glueball state may be obtained by taking the ordered product of $SU(N)$ link matrices around a contractible loop and then taking the trace. The real(imaginary) part of the trace projects onto charge conjugation $C = + (-)$. We sum all spatial translations of the loop so as to obtain an operator with zero momentum. We take all rotations of the loop and construct the linear combinations that transform according to the irreducible representations, $R$, of the rotational symmetry group of our cubic spatial lattice. We always choose to use a cubic spatial lattice volume ($L_x=L_y=L_z$) that respects these symmetries. For each loop we also construct its parity inverse so that taking linear combinations we can construct operators of both parities, $P = \pm$. The correlators of such operators will project onto glueballs with momentum $p = 0$ and the $R^{P C}$ quantum numbers of the operators concerned. A number of the paths used for the construction of our basis are provided in Figure~\ref{fig:glueball_operators}.
The irreducible representations $R$ of our subgroup of the full rotation group are
labelled as $A_1,A_2,E,T_1,T_2$. The $A_1$ is a singlet and rotationally symmetric, so it
will contain the $J=0$ state in the continuum limit. The $A_2$ is also a singlet, while the
$E$ is a doublet and $T_1$ and $T_2$ are both triplets. Since, for example, the three states transforming as the triplet of $T_2$ are degenerate on the lattice, we average their values and treat them as one state in our estimates of glueball masses and we do the same with the $E$ doublets.

The glueball energy states are extracted using correlation matrices $C_{ij} = \langle \phi_i^{\dagger} (t) \phi_j (0) \rangle$ with $i,j=1...N_{\rm op}$ in combination with GEVP where $N_{\rm op}$ is the number of operators. The scalar channel $A_1^{++}$ has a non-zero projection onto the vacuum and we, thus, use the vacuum-subtracted operator $\phi_i - \langle \phi_i \rangle$, to remove the contribution of the vacuum in Eq.~\ref{extract_mass}, so that the lightest non-trivial state is the leading term in the expansion of states.


The above representations of the rotational symmetry reflect our cubic lattice formulation. As we approach the continuum, these states will approach the continuum glueball states which belong to representations of the continuum rotational symmetry. In other words they fall into degenerate multiplets of $2J + 1$ states. In determining the continuum limit of the low-lying glueball spectrum, it is clearly more useful to be able to assign the states to a given spin $J$, rather than to the representations of the cubic subgroup which have a much less fine ‘resolution’ since all of $J =1,2,3 \dots, \infty$,  are mapped to just 5 cubic representations. The way the $2J + 1$ states for a given $J$ are distributed amongst the representations of the cubic symmetry subgroup is given, for low values of $J$, in Tab.~\ref{tab:table_J_R}. 
\begin{table}[ht]
\begin{center}
 \begin{tabular}[h]{|c|ccccc|}
\hline
    $J$ & $A_1$ & $A_2$ & $E$ & $T_1$ & $T_2$ \\
    \hline
    0 & 1 & 0 & 0 & 0 & 0 \\
    1 & 0 & 0 & 0 & 1 & 0 \\
    2 & 0 & 0 & 1 & 0 & 1 \\
    3 & 0 & 1 & 0 & 1 & 1 \\
    4 & 1 & 0 & 1 & 1 & 1 \\
\hline
  \end{tabular}
\end{center}
\caption{Subduced representations of continuum spin $J \downarrow R$ of the octahedral group of rotations up to $J=4$, illustrating the spin content of the
   representations $R$ in terms of the continuum $J$.}
\label{tab:table_J_R}
\vspace{-0.5cm}
\end{table}
\vspace{-0.05cm}

\section{Topological charge and scale setting}
\label{sec:topological_charge_and_scale_setting}
\vspace{-0.25cm}
{\color{black}
 In the continuum, the topological charge is defined as the integral over the four-dimensional volume of the topological charge density namely
\be
{Q} = \frac{1}{32\pi^2} \int d^4 x \: \epsilon_{\mu\nu\rho\sigma} \Tr\left[F_{\mu\nu}(x)F_{\rho\sigma}(x)\right] \,.
\label{eq:Q_continuum_def}
\ee
The discrete counterpart of the above quantity can be obtained by replacing the gluonic field tensor with a lattice operator that reproduces the correct continuum limit. The choice is not unique, and operators with smaller discretization effects can be obtained by using \({\cal O}(a)\)-improved definitions of \(F_{\mu\nu}\). The definition of \({Q}\) we choose to use in this work is the symmetric definition, first introduced in Ref.~\cite{DiVecchia:1981aev}. 
We use the gradient flow \cite{Luscher:2010iy} in order to suppress the UV fluctuations of the gauge field defining the topological charge. 
The smoothing action employed in the flow equation is the standard Wilson action. The elementary integration step is \(\epsilon=0.01\) and the topological charge is computed on the smoothed fields at multiples of \(\Delta \tau_{\rm flow}=0.1\).
The flow time must be chosen large enough such that discretization effects are negligible but not so large that the topological properties of the gauge field are changed; this happen at \(a\sqrt{8\tau_{\rm flow}}\sim \mathcal{O}(0.1\text{fm})\) according to Ref.~\cite{Luscher:2010iy}.
%


The gradient flow also enables the definition of a physical scale parameter $t_0$, which can be determined to high precision.  This flow observable was introduced in~\cite{Luscher:2009eq,Borsanyi:2012zs}. $t_0$ is defined according to the following prescription. First, we set
    \begin{eqnarray}
        F(t) = t^2 \langle E(t) \rangle \, \ {\rm where} \ E(t) = \frac{1}{4} B^2_{\mu \nu} (t)\,,
    \end{eqnarray}
    where $B_{\mu \nu}$ is field strength obtained by flowing $F_{\mu \nu}$ along the flow time direction. We define the scale $t_0(c)$ as the value of $t$ for which $F(t) |_{t=t_0(c)} = c\,$ where $c$ should be chosen so that the relevant condition $a \ll \sqrt{8 t_0} \ll L$ is satisfied. Small $c$ leads to larger lattice artefacts and larger $c$ usually lead to larger autocorrelations \cite{Bergner:2014ska}. In our case we choose the value $c=0.3$ which is the value commonly used in lattice QCD calculations.

    In Fig.~\ref{fig:topological_charge} we present the history of the topological charge as well as its distribution for our reference ensemble \texttt{cB4.06.24}. Clearly the plots do not indicate topological freezing, suggesting that the Markov-Chain is ergodic. Furthermore the distribution of the topological charge is to a good extent Gaussian-shaped suggesting our calculations have correctly probed the topological sectors of the theory. All other ensembles exhibit similar behaviour.
    \vspace{-0.5cm}
\begin{figure}[h]
    \centering
    \includegraphics[height=3.75cm]{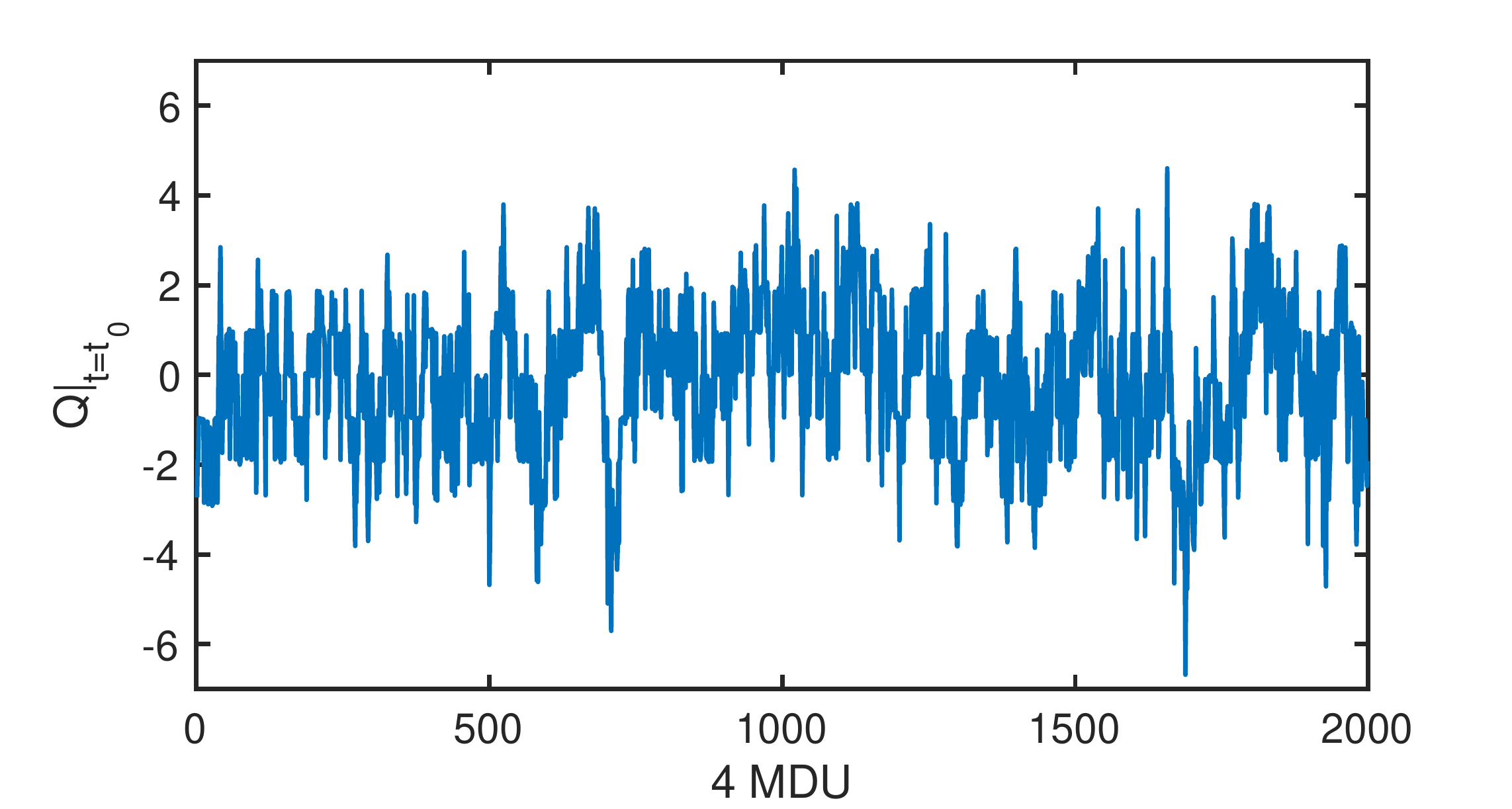}
    \includegraphics[height=3.75cm]{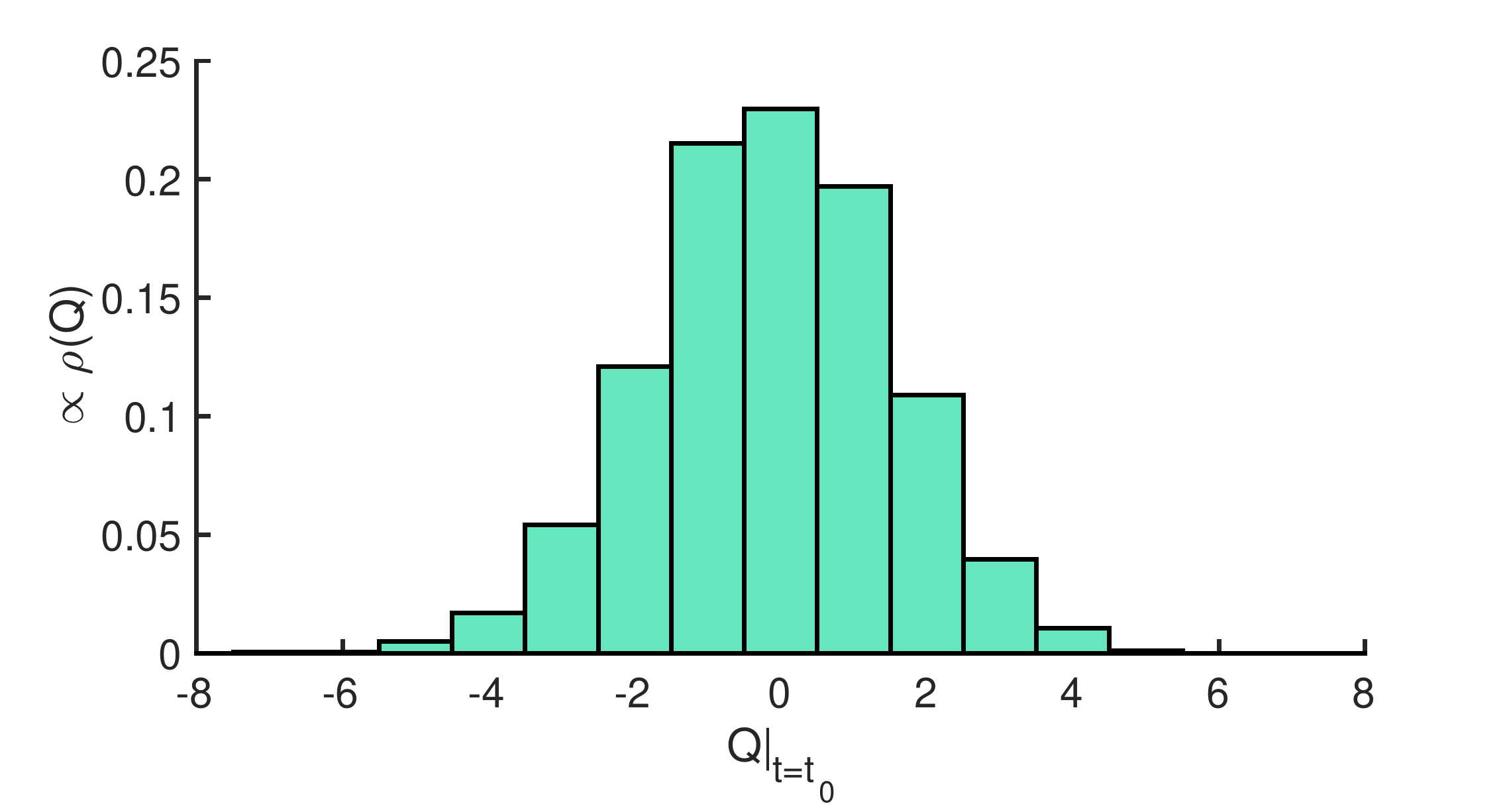}
    \caption{The history of the topological charge as well as its distribution for \texttt{cB4.06.24} at $t=t_0$.}
    \label{fig:topological_charge}
\end{figure}
\vspace{-0.5cm}
\section{Results}
\label{sec:results}
\vspace{-0.25cm}
We have extracted the low lying spectrum for the irreducible representations $A_1^{++}$, $E^{++}$, $T_2^{++}$ as well as $A_1^{-+}$, corresponding to the scalar, tensor as well as to the pseudoscalar channels. The first striking feature of the calculation is that the effective mass plateaus (Fig.~\ref{fig:plots_effective_masses_glueballs}) set in relatively late in time, leading to an overlap of 30 - 50 \%   which compares badly with the fast converging mass plateaus one encounters in the pure gauge theory with overlaps of 90 - 100 \%. This indicates that the Hilbert space of the tower of states is rather rich due to the inclusion of dynamical quarks and, thus, of the mixings with mesonic states. We have attempted to increase the overlaps onto the extracted states by increasing the variational basis of operators, however, with negligible differences.
\begin{figure}[h]
    \vspace{-0.5cm}
    \centering
    \includegraphics[height=4.2cm]{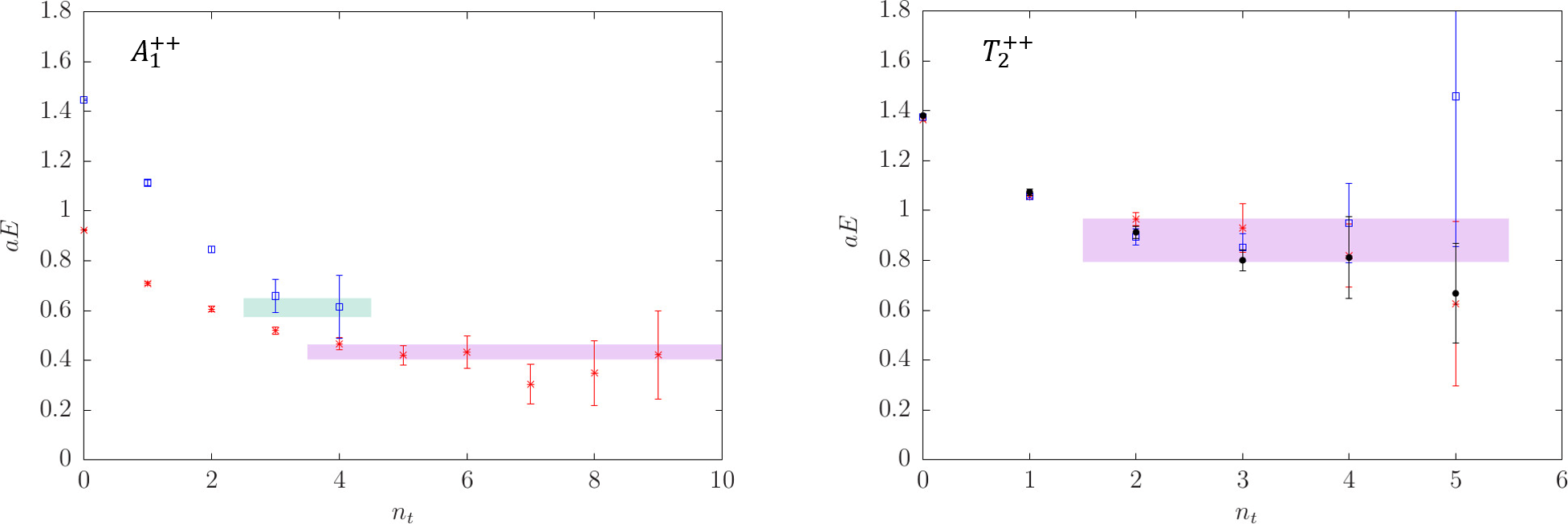} 
    \caption{ \underline{Left panel:} The effective masses for the ground and first excited states of the irreducible representation $A_1^{++}$ for the ensemble \texttt{cB4.06.24}. \underline{Right panel:} The effective mass for the ground state of the irreducible representation $T_2^{++}$ for the ensemble \texttt{cB4.06.24}; states in $T_2$ representation come in triplets, thus, the three states that appear as ground, first excited and second excited states within the variational analysis.}
    \label{fig:plots_effective_masses_glueballs}
     \vspace{-0.05cm}
\end{figure}

In the left, middle and right panels of Fig.~\ref{fig:plots_masses_glueballs} we present the results for $N_f=4$, $\{\beta=1.778 , L=16\}$, $\{\beta=1.778 , L=24\}$ and $\{\beta = 1.836, L = 24\}$ respectively as well as the comparison with the pure gauge theory. We consider the plot in the middle of Fig.~\ref{fig:plots_masses_glueballs} as our reference plot since we know that the pion mass does not experience finite volume effects in this case. (Glueball masses appear to be insensitive to finite volume effects.) This can be seen by comparing the plot in the middle where the spatial lattice size is $L=24$ with the one on the left where $L=16$ while the value of $\beta$ is kept fixed. Nevertheless, lowering the volume can give rise to other non-glueball states such as di-torelons. As a matter of fact on the left panel, for the representation $E^{++}$ the ground state is a di-torelon with mass slightly larger than twice the mass of the torelon. As expected by increasing the lattice spatial length $(L=16 \to 24)$, this state effectively disappears. In addition we investigated whether a reduction in the lattice spacing would affect the spectrum of $N_f=4$, demonstrating that, at least for the low lying spectrum, the spectrum presented in the middle plot of  Fig.~\ref{fig:plots_masses_glueballs} does reflect continuum physics. This can be seen by comparing the spectrum on the right of Fig.~\ref{fig:plots_masses_glueballs} where we alter the value of $\beta$.

\begin{figure}[h]
\vspace{-0.5cm}
    \centering
    \scalebox{1}{\includegraphics[height=7.7cm]{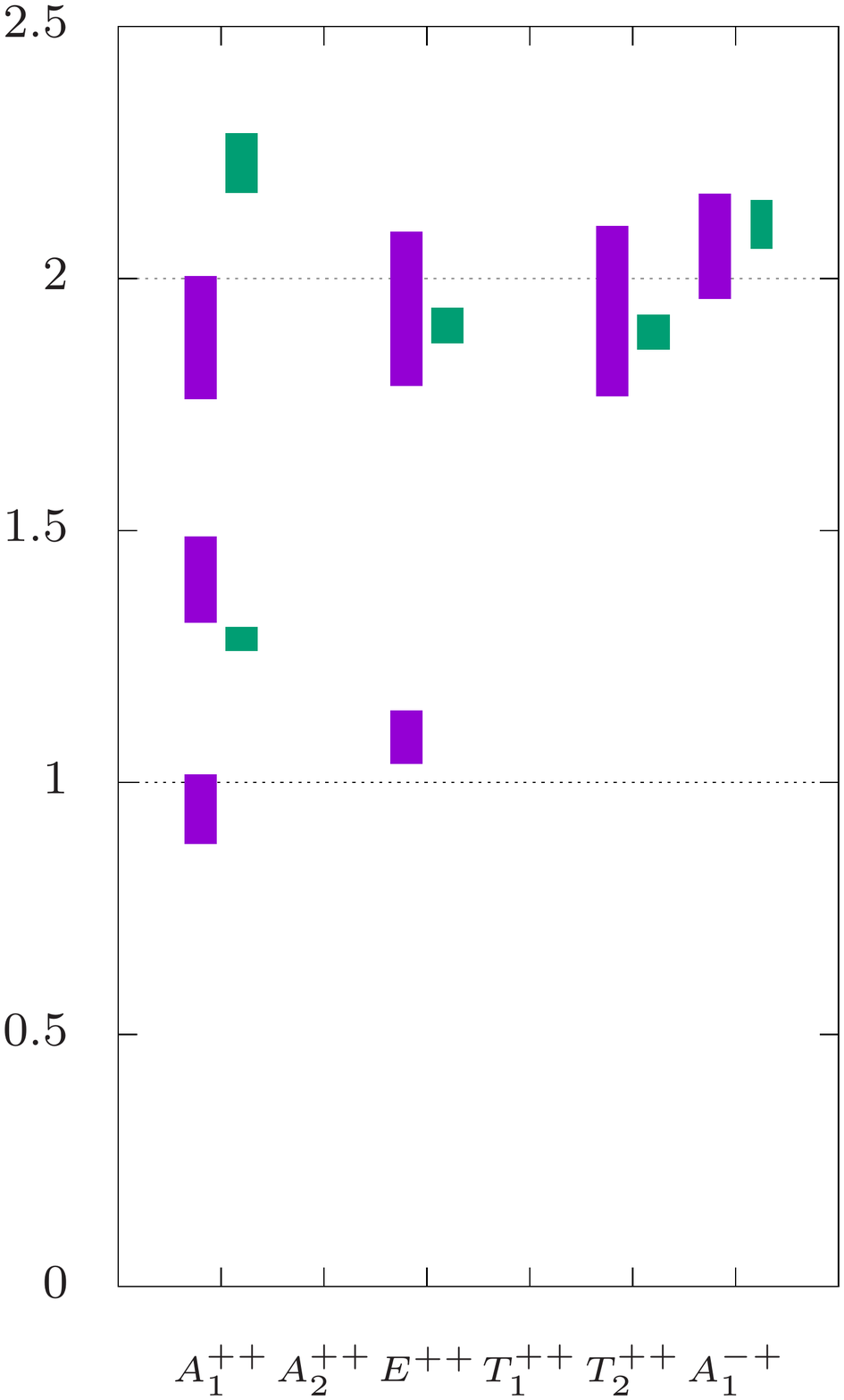} \hspace{-1.5cm}\includegraphics[height=7.7cm]{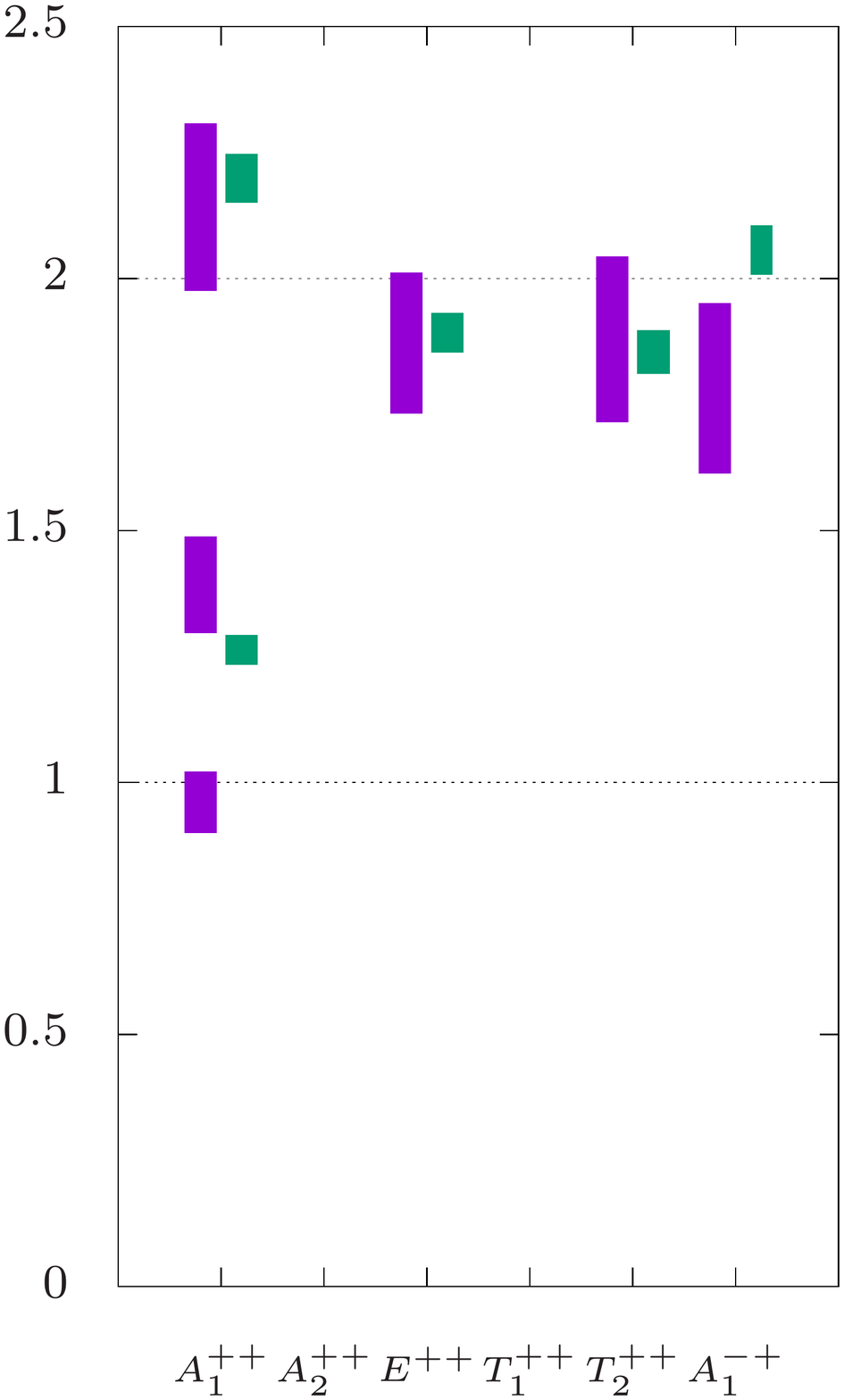}
    \hspace{-1.5cm}\includegraphics[height=7.7cm]{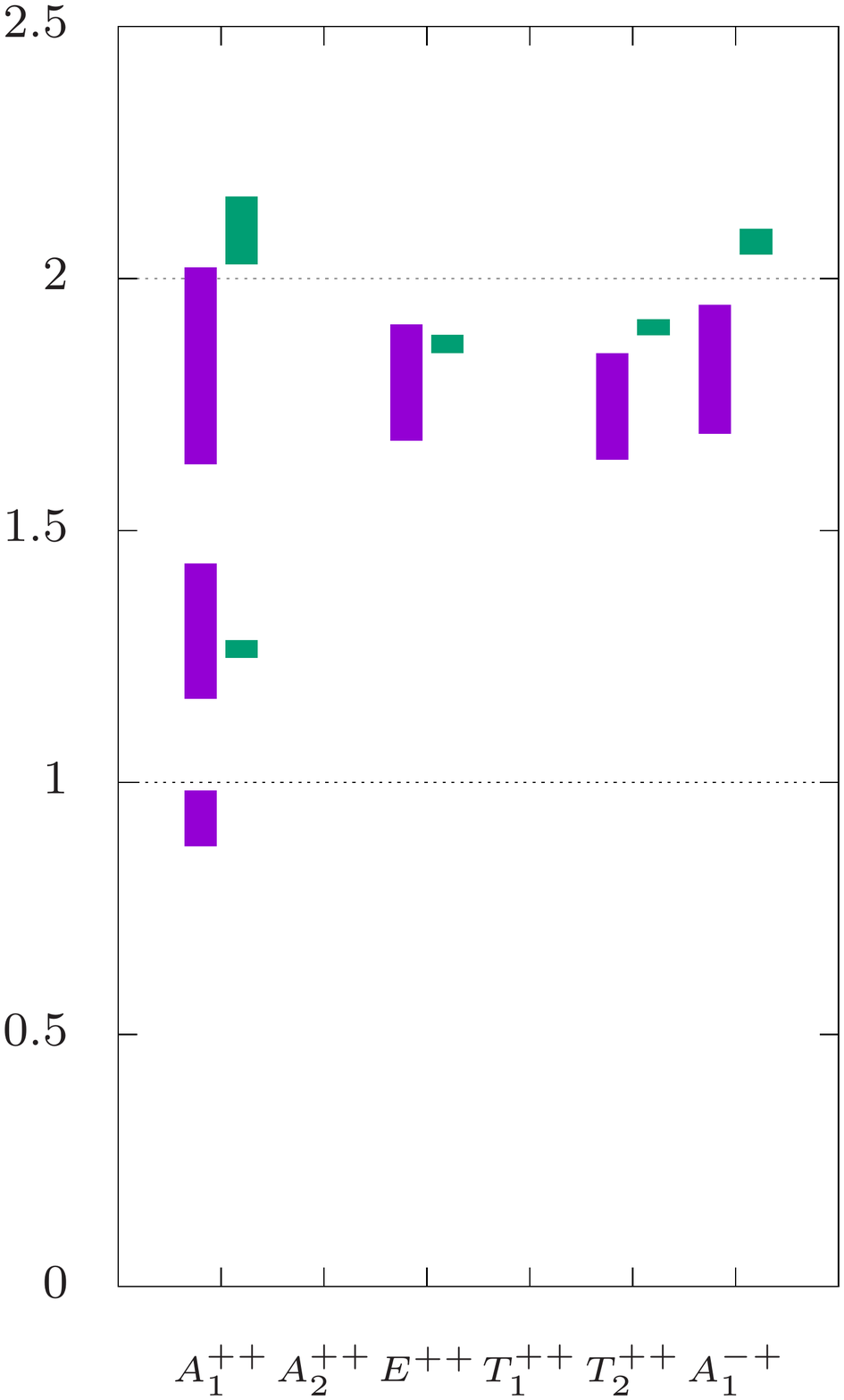} \put(-428,110){\scriptsize $\sqrt{t_0} M_{R^{PC}}$}}
    \caption{The spectrum of glueballs for the representations $A_1^{++}$, $E^{++}$, $T_2^{++}$, $A_1^{-+}$ for the ensembles \texttt{cB4.06.16}(left panel), \texttt{cB4.06.24}(middle panel) 
    and \texttt{cC4.05.24}(right panel). The states of $N_f=4$ QCD are denoted in purple while the states of pure gauge $SU(3)$ in green.}
    \label{fig:plots_masses_glueballs}
    \vspace{-0.05cm}
\end{figure}
Turning now to the comparison between $N_f=4$ QCD and the $SU(3)$ pure gauge theory we first extract the value of $t_0/a^2$ for $N_f=4$ configurations and we subsequently extract the spectrum of glueballs for the pure gauge theory at a value of $\beta$ that corresponds to the same $t_0/a^2$. The values of $t_0/a^2$ are given in Table~\ref{tab:params_sim_Nf2+1+1} while the values of $\beta$ have been extracted using interpolations of data in Refs~\cite{Luscher:2010iy,Francis:2015lha,Luscher:2011kk,Ce:2015qha} and presented in Table~\ref{tab:params_sim_pure_gauge}. Since $t_0/a^2$ is defined using purely gluonic variables, just like our glueball operators, and since it is much more precise than the masses, comparing dimensionless ratios $M_G\surd{t_0}$ in this way, between the two different theories, is a plausible, although not unique, strategy.

The glueball masses for pure gauge theory are represented in green color while the masses for $N_f=4$ are in purple. Strikingly, with $N_f=4$ the ground state within the $A_1^{++}$ tower of states is a state that is additional to the states in the pure gauge theory. In contrast we observe an adequate agreement of the pure gauge $A_1^{++}$ ground state with the first excited $A_1^{++}$ state of $N_f=4$. This immediately leads to the assumption that the ground state of $N_f=4$ includes significant quark content while the first excited state corresponds to the actual glueball state. Needless to say this scenario requires further investigation. 

Turning now to the ground state of $E^{++}$ and $T_2^{++}$ and, thus, to the ground state of $2^{++}$, it appears that there is an agreement between $N_f=4$ and the pure gauge theory, suggesting that accurate pure gauge calculations of the tensor glueball can be used for comparison with experiment. Finally, a similar situation obtains for the $A_1^{-+}$ pseudoscalar ground state, although there is some suggestion, at the level of one sigma, that it becomes slightly lighter when dynamical quarks are inserted into the simulation. Furthermore, in $N_f=4$ the ground state of $2^{++}$ is close to that of the $0^{-+}$, just as in the pure gauge theory.
\vspace{-0.5cm}

\section{The $A_1^{++}$ ground state for $N_f=2+1+1$}
\label{sec:Nf211}
\vspace{-0.25cm}
To further investigate the nature of the additional $A_1^{++}$ state seen in Fig.~\ref{fig:plots_masses_glueballs} we carried out a simulation in $N_f=2+1+1$ QCD. To be specific, we extracted the low-lying $A_1^{++}$ spectrum in the more realistic case of $N_f=2+1+1$ twisted mass fermions for two values of the pion mass, $m_{\pi} \sim 260$ and $\sim 350$ MeV. Once more we have made use of ensembles consisting of as many configurations as possible. Hence, for the purposes of this investigation we found suitable the choice of the ensembles {\texttt{cA211.53.24}} (5000 confs) and \texttt{cA211.25.32} (2500 confs) the details of which are presented in Table~\ref{tab:params_sim_Nf2+1+1}.
\begin{figure}[h]
\vspace{-0.25cm}
    \centering
    \rotatebox{0}{\includegraphics[height=6.0cm]{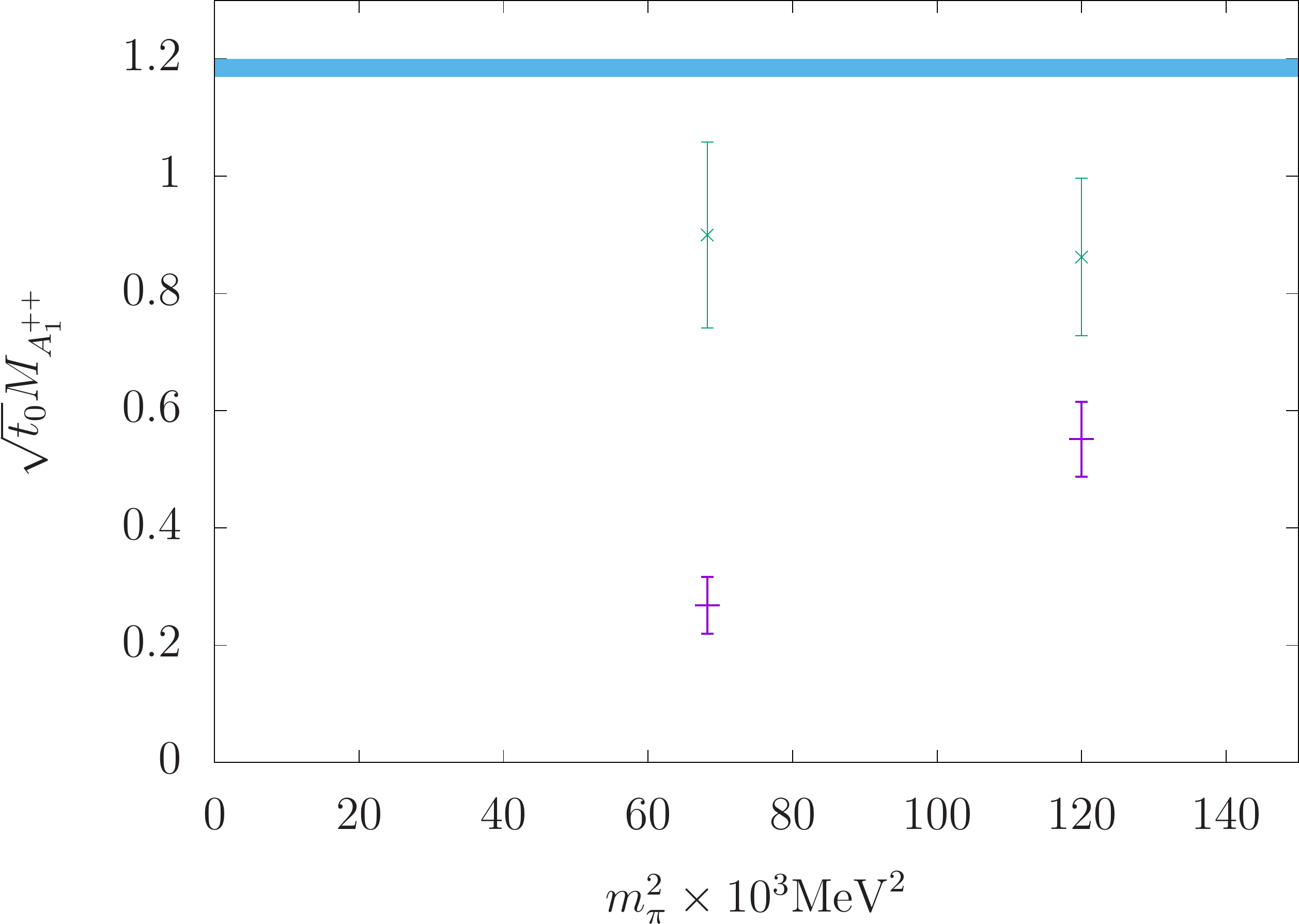}}
    \caption{The ground state as well as the first excited state of the scalar channel $A_1^{++}$ for the ensembles \texttt{cA211.53.24}  and \texttt{cA211.25.32}. The blue band represents the scalar glueball mass of the $SU(3)$ pure gauge theory at $\beta=5.8941$ $(t_0/a^2 \sim 2.2)$ taken from Ref.~\cite{Athenodorou:2021qvs}.}
    \label{fig:spectrum_Nf211}
\end{figure}


The results, presented in Fig.~\ref{fig:spectrum_Nf211} show the ground as well as the first excited state for the two different pion masses. The message is unambiguous, namely the ground state depends strongly on the pion mass while the first excited state is approximately constant. This leaves little doubt that the ground state has a dominant quark content, possibly representing a decay of a glueball to either two or four pions. We note that the mass of the ground state is to a good extent close to the mass of two pion masses. The first excited state which appears to  be roughly constant fits the scenario of representing the lightest glueball. Of course the lattice is coarse and, thus, we expect the masses to be significantly influenced by lattice artifacts; hence, presumably, the significant deviation from the pure gauge value.

 \section{Conclusions}
\label{sec:conclusions}
\vspace{-0.25cm}
In this work we have investigated the spectrum of glueball masses for the scalar, tensor and pseudo-scalar channels, $J=0^{++}$, $2^{++}$ and $0^{-+}$ respectively, in the presence of dynamical quarks. We have first extracted the spectrum of glueballs on configurations produced with $N_f=4$ light quarks with $m_\pi \sim 250$ MeV. This includes an investigation of possible finite volume and discretisation effects. We adopt as our point of reference the \texttt{cB4.06.24} ensemble for which the pion mass exhibits negligible finite volume effects. We compare the resulting masses with the corresponding pure gauge configurations using $t_0/a^2$ as the physical scale, and we do so at comparable values of  $t_0/a^2$ so as to hopefully minimise any distortion from lattice corrections (which in any case are likely to be small compared to the large statistical uncertainties). Strikingly, the $A_1^{++}$ spectrum appears to include an additional state, which for our light quark masses happens to be the lightest state. The fact that the first and second excitation levels of the $A_1^{++}$, $N_f=4$ spectrum have masses close to the ground and first excited states of the pure gauge theory, suggests that this $A_1^{++}$, $N_f=4$  ground state is a multi-pion or quarkonium state.  Furthermore, the $2^{++}$, $N_f=4$ ground state is in a good agreement with the pure gauge $2^{++}$ ground state. In addition, the pseudoscalar $A_1^{-+}$, $N_f=4$ ground state is close to the pure gauge $A_1^{-+}$ ground state. To shed more light on the additional state appearing in the $A_1^{++}$ channel we extracted the spectrum of the $A_1^{++}$ with $N_f=2+1+1$ for a fixed lattice spacing of $\sim 0.09$ fm and two pion masses revealing that the mass of the $A_1^{++}$ depends strongly on the pion mass. This is a clear indication that the $A_1^{++}$ additional state has a large "quark content" and is not a glueball, something that needs to be addressed in more detail in future investigations. Finally, our findings in combination with data obtained from other projects suggest that the actual glueball spectrum depends negligibly on the mass of dynamical quarks.

\section{Acknowledgements}
\vspace{-0.25cm}
A.A. is supported by the Horizon 2020 European research infrastructures programme "NI4OS-Europe" with grant agreement no. 857645. Results were obtained using Cyclone High Performance Computer at The Cyprus Institute, under access with id \texttt{p014}, the Oxford Theoretical Physics cluster, the resources of computer cluster Isabella based in SRCE - University of Zagreb University Computing Centre, as well as using computing
time granted by the John von Neumann Institute for Computing (NIC) on the supercomputer
 JUWELS Cluster~\cite{JUWELS} at the J\"ulich Supercomputing Centre (JSC).
J.F.~is financially supported by the H2020 project PRACE 6-IP (GA No. 82376) and by the EuroCC project (GA No. 951732) funded by the Deputy Ministry of Research, Innovation and Digital Policy and the Cyprus Research and Innovation Foundation and the European High-Performance Computing Joint Undertaking (JU) under grant agreement No 951732.

\vspace*{-0.3cm}
\bibliographystyle{JHEP}
\bibliography{biblio_NEW}
\end{document}